\documentclass[]{amsart}

\usepackage[style=nature, backend=biber]{biblatex}

\usepackage{graphicx,amsfonts}% Include figure files

\newtheorem{theorem}{Theorem}

\newtheorem{remark}{Remark}

\newcommand{\bee}{\begin{eqnarray}}
\newcommand{\eee}{\end{eqnarray}}
\newcommand{\be}{\begin{eqnarray*}}
\newcommand{\ee}{\end{eqnarray*}}
\newcommand{\R}{{\mathbb R}}
\newcommand{\N}{{\mathbb N}}
\newcommand{\C}{{\mathbb C}}
\newcommand{\I}{{\mathbb I}}

\newcommand{\D}{{\mathcal D}}

\newcommand{\B}{{\mathcal B}}
\newcommand{\U}{{\mathcal U}}
\newcommand{\A}{{\mathcal A}}
\newcommand{\vvr}{{\mathbf x}}
\newcommand{\vR}{{\mathbf R}}
\newcommand{\vk}{{\mathbf k}}
\newcommand{\vG}{{\mathbf G}}
\newcommand{\w}{w}
\newcommand{\E}{\mathcal E}

\begin{document}

\title [Resonance crossings, diabolical points, and quantum sensors]{Resonance crossings and diabolical points for DC electric field quantum sensors}

\author {Andrea Sacchetti}

\address {Department of Physics, Informatics and Mathematics, University of Modena e Reggio Emilia, Modena, Italy}

\email {andrea.sacchetti@unimore.it}

\date {\today}

\begin {abstract}
Since the 1990s, experiments on double-well heterostructures have highlighted a key phenomenon: applying an external DC electric field induces pronounced wave packet oscillations at specific field strengths. \ To gain a deeper insight into this effect, a comprehensive quantitative analysis of the properties of quantum resonances near a crossing point is required. \ This approach lays the theoretical foundation for devising a quantum sensor capable of rapidly and efficiently determining whether the strength of an external DC electric field lies within a narrow target range.
\end{abstract}


\maketitle

\section {Introduction} For time-independent  Schr\"odinger equations with a parameter-dependent double-well potential, the eigenvalues typically exhibit level repulsion as the parameters vary. \ This behavior changes significantly upon applying an external DC electric field, and our attention must shift to the behavior of complex-valued quantum resonances. \ Indeed, unlike real eigenvalues, two quantum resonances can cross at critical parameter values, giving rise to one of two resonance crossings (RC) scenarios \cite{Avron1982,Grecchi1995,Korsch2003}:
\begin{itemize}

\item[-] Type~$I$ RC, where the real parts of the two quantum resonances exhibit an avoided crossing pattern, akin to that observed for real eigenvalues, while the imaginary parts cross;

\item[-] Type~$II$ RC, where the imaginary parts of the two quantum resonances exhibit an avoided crossing pattern while the real parts cross.

\end{itemize}
In the literature, the degenerate case corresponding to an exact intersection is referred to as a \emph{diabolical point} \cite{Berry1984}, or also an exceptional point. \ To analyze the behavior of two quantum resonances in a neighborhood of a crossing point, the Schr\"odinger operator with a tilted double-well potential can be reduced, by selecting a suitable orthonormal basis, to a non-symmetric $2\times 2$ matrix whose eigenvalues coincide with the quantum resonances and whose associated eigenvectors represent the corresponding metastable states. \ It is essential to emphasize that for a Type~$I$ RC, the two aforementioned metastable states are equally localized in both potential wells; whereas for a Type~$II$ RC, each metastable state is localized within a single, distinct well  \cite{Heiss2012,Heiss2016}. \ This feature plays a crucial role in qualitatively explaining the time evolution of physical observables. \ Specifically, in the presence of a Type~$I$ RC, a periodic beating dynamics of the wavefunction between the two wells is observed, whereas for a Type~$II$ RC, such beating dynamics is not expected to occur. \ Hence, diabolical points are critical, as they delineate the boundaries between parameter regimes associated with qualitatively distinct quantum dynamics.

\begin{figure}
\begin{center}
\includegraphics[height=5cm,width=8.5cm]{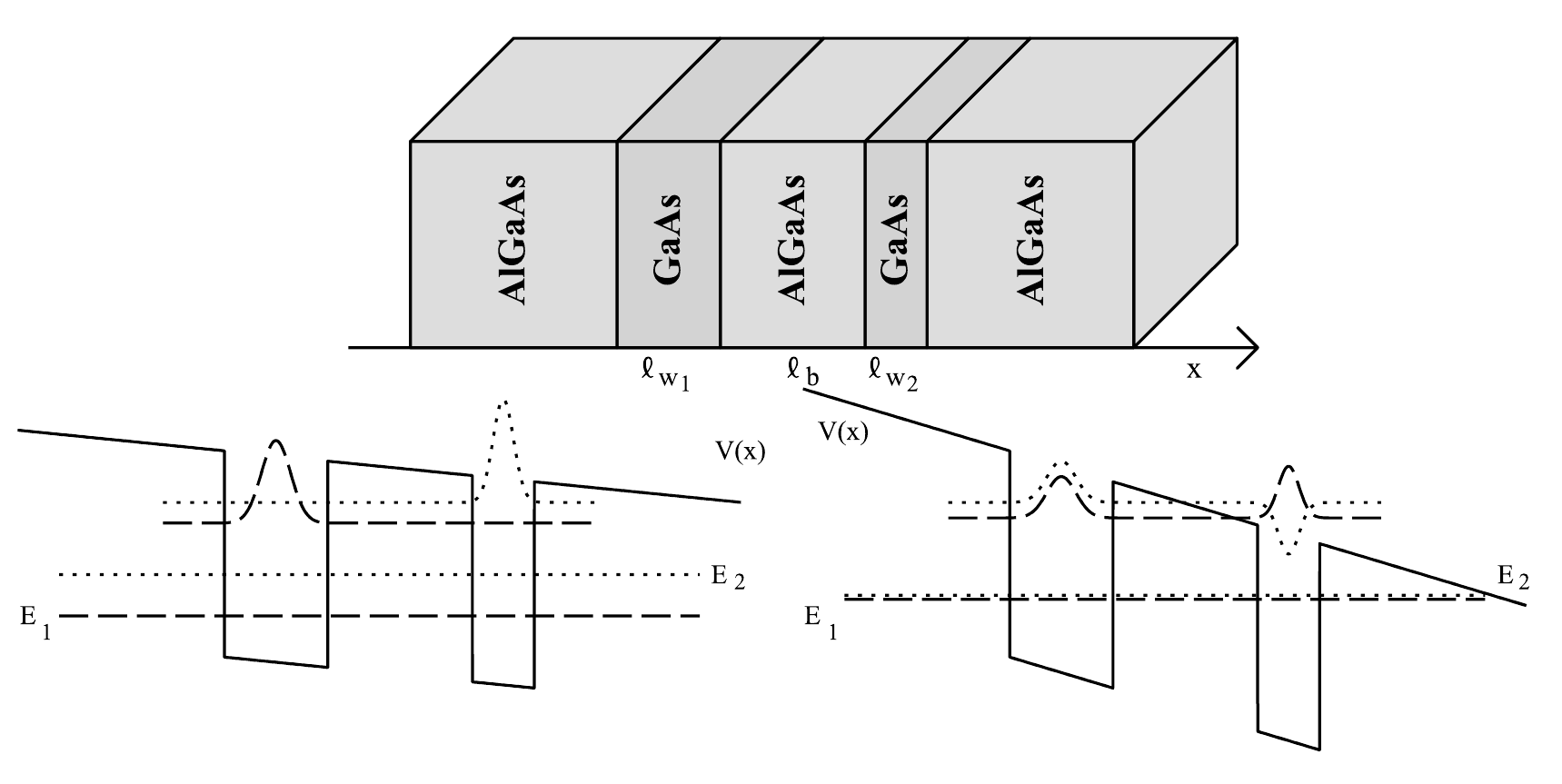}
\caption{Schematic representation of the heterostructure (\ref{F1}) under an applied DC electric field along the growth direction $x$. \ Within the envelope-function approximation, the potential $V(x)$ corresponds to a tilted double square well with no bound states, supporting instead the quantum resonances $E_1$ (solid line) and $E_2$ (dashed line).}
\label {Fig1}
\end{center}
\end{figure}

This theoretical framework can find practical application in real-world devices. \ More specifically, we consider experimental observations of quantum beating in heterostructures subjected to a DC electric field, a phenomenon whose investigation was initiated in the early 1990s by the pioneering studies of Leo {\it et al.}~\cite{Leo1991} and Roskos {\it et al.}~\cite{Roskos1992}; see also \cite{Heberle1994} and, for reviews, \cite{Ferreira1997,Leo2011}. \ Following the above-mentioned experiments, we focus on the case of two $\mathrm{GaAs}$ layers embedded between three $\mathrm{Al}_s\mathrm{Ga}_{1-s}\mathrm{As}$ ($s\in (0,1)$) layers:
%
%\begin{widetext}
\bee
\mathrm{Al}_s\mathrm{Ga}_{1-s}\mathrm{As} / \ell_{w_1} \cdot \mathrm{GaAs} / \ell_{b_1}\cdot \mathrm{Al}_s\mathrm{Ga}_{1-s}\mathrm{As} / \ell_{w_2} \cdot \mathrm{GaAs} / \mathrm{Al}_s\mathrm{Ga}_{1-s}\mathrm{As} \label{F1}
\eee
%\end{widetext}
%
where $\ell_{w_1}$ and $\ell_{w_2}$ denote the widths of the two $\mathrm{GaAs}$ quantum wells, and $\ell_{b_1}$ is the width of the internal $\mathrm{Al}_s\mathrm{Ga}_{1-s}\mathrm{As}$ barrier. \ We then assume that the device is subjected to a DC electric field applied along the growth direction $x$ of the heterostructure (i.e. perpendicular to the layers; see Fig.~\ref{Fig1} - top panel).

Upon applying the DC electric field, the system does not admit bound states, but rather quantum resonances. \ When the energies of two adjacent resonances, $E_1$ and $E_2$, are sufficiently separated (note that $E_1$ and $E_2$ denote here a generic pair of adjacent interacting resonances, rather than the ground and first excited states), each metastable state is localized in just one well (see Fig.~\ref{Fig1} - bottom left-hand panel). \  However, since the Stark potential contributes differently to the left and right wells the two energies become nearly aligned  for a critical value of the electric field strength. \ Under this energy-alignment condition, provided that the RC is of Type~$I$, resonant tunneling causes the associated metastable states to become equally localized across both wells (see Fig.~\ref{Fig1} - bottom right-hand panel) and thus an interference effect is triggered and large-amplitude periodic oscillations of some physical observable occurs.

In this work, going beyond this qualitative explanation, we introduce precise quantitative techniques for realistic models with non-constant effective masses: by analyzing the behavior of quantum resonances near a crossing point, we provide an analytical explanation of the experimental findings, achieving excellent agreement. \ Furthermore, we are able to predict the beating dynamics of the wavefunction as the model parameters vary; in this way, one can tune the heterostructure parameters to conceive a quantum system that exhibits pronounced oscillations only when the external electric field strength assumes precise, predefined values. \ Hence, a novel quantum sensing scheme is devised for the rapid, straightforward, and highly precise verification of whether the strength of the external DC electric field matches a target value.

It is also worth highlighting several earlier theoretical studies on double-well heterostructures. \ For instance, see \cite{Pedersen} for an analytical investigation of bound-state behavior under an external confining potential as a function of the electric field strength, and \cite{Sival} for a detailed analysis of resonance states in systems with $s=0.33$. \ More recently, a toy model in which the double-well potential is represented by a pair of attractive Dirac $\delta$-potentials was employed \cite{Sacchetti2024}. \ Finally, see \cite{Sacchettia2026}  where constant effective masses were assumed within the envelope-function approximation (\ref{F2}).

The paper is organized as follows. \ In Section \ref {Sec2}, we summarize the experimental results obtained by Roskos {\it et al.} \cite {Roskos1992} and find excellent agreement when comparing them with the predictions of the theoretical model. \ In Section \ref {Sec3}, we show that tuning the theoretical model parameters appropriately allows for a single Type~$I$ RC, significantly narrowing the range over which beating dynamics are observed. \ Finally, in Section \ref {Sec4}, we revisit the theoretical model to devise a quantum sensor capable of quickly and easily determining whether the electric field applied to the heterostructure matches a predefined value with high accuracy, within a good degree of tolerance. \ Finally: in Appendix \ref{App1}, we collect some known results concerning the envelope function approximation to make the paper self-contained; in Appendix \ref{App2}, we prove that the Schrödinger operator (\ref{F2}) admits a unique self-adjoint extension, this is a technical result that allows us to express the solution to the time-dependent Schr\"odinger equation by means of the evolution operator; in Appendix \ref{App3}, we employ the exterior complex scaling method to derive a formula for computing the survival amplitude in terms of the metastable states associated with quantum resonances.

\section {Method: theoretical analysis of experiments} \label {Sec2} In the experiment reported by \cite{Roskos1992}, where a relationship between the electric field strength and the wavefunction beat dynamics has been thoroughly measured, the double-well device was fabricated from a heterostructure comprising two $\mathrm{GaAs}$ wells with widths $\ell_{w_1}
%= 145\ \text{\AA}
$ and $\ell_{w_2}
%= 100\ \text{\AA}
$, separated by an $\mathrm{Al}_{s}\mathrm{Ga}_{1-s}\mathrm{As}$ barrier of width $\ell_{b_1}
%= 25\ \text{\AA}$, where $s=0.2
$. \ The wavefunction was initially prepared in the wider well; upon applying an external DC electric field, dipole radiation was observed, originating from oscillations of the wavefunction at frequencies of the order of $1.5\ \mathrm{THz} $. \ More precisely, these oscillations exhibited a significant amplitude for DC electric field strengths exceeding $6.7\ \mathrm{kV/cm}$, whereas for field strengths of $15.8\ \mathrm{kV/cm}$ or higher, the oscillation amplitudes decayed so rapidly that their frequency could not be determined.

In the envelope-function approximation, the dynamics of the quantum system is described by means of the one-dimensional Schr\"odinger operator formally defined on $L^2 (\mathbb{R})$ as (see, e.g., Sec.~2.1 in \cite{Altarelli1986}):
%
%\begin{widetext}
\bee
H \psi =
- \frac{\hbar^2}{2m_j} \frac{\mathrm{d}^2 \psi}{\mathrm{d} x^2} + [v_j - e F x] \psi\, , \ x_{j-1} < x < x_j\, ,  \label{F2}
\eee
%\end{widetext}
%
where $j=1,\ldots ,5$, $x_0 = -\infty$, $x_1 = 0$, $x_2 = \ell_{w_1}$, $x_3 = x_2 + \ell_{b_1}$, $x_4 = x_3 + \ell_{w_2}$, and $x_5 = +\infty$, and the time-dependent Schr\"odinger equation takes the form
\bee
\left\{
\begin{array}{l}
i \hbar \frac{\partial \psi}{\partial t} = H \psi \, , \\[1.5ex]
\left. \psi(x,t) \right|_{t=0} = \psi_0(x) \, ,
\end{array}
\right.
\label{F3}
\eee
where $\psi_0(x)$ is the initial wavefunction; $\psi (x,t)$ is the envelope function. \ Here, $m_1 = m_3 = m_5 := m_b^\star$ and $m_2 = m_4 := m_w^\star$ are the effective masses of the electron in the $\mathrm{Al}_s\mathrm{Ga}_{1-s}\mathrm{As}$ and $\mathrm{GaAs}$ semiconductors, respectively; $e$ is the elementary charge; $\mathcal{E}$ is the strength of the electric field; and $F := -\mathcal{E}$ (still named, when this fact does not cause misunderstanding, electric field strength). \ Furthermore, $v_1 = v_3 = v_5 = 0$ and $v_2 = v_4 := v$, with $v$ being the conduction-band offset between the two semiconductors. \ Since $v < 0$, this yields a double square well structure with well widths $\ell_{w_1}$ and $\ell_{w_2}$, separated by an inner barrier of width $\ell_{b_1}$. \ The effective masses are given by $m_w^\star = c_w m_0$ and $m_b^\star = c_b m_0$, where $m_0$ is the free-electron mass and $c_w = 0.067$ for the $\mathrm{GaAs}$ lattice. \ We point out that both $c_b$ and $v$ depend on $s$; for concreteness, in our analysis we adopt the following linear relations \cite{adachi1993,vurgaftman2001}:
\bee
v(s) \approx -810 \, s \ \mathrm{meV} \quad \text{and} \quad c_b(s) \approx 0.067 + 0.083 \, s \, , \label{F4}
\eee
in the range $s \in [0, 0.45]$ at room temperature ($T = 300\ \mathrm{K}$).

To compare the results reported in \cite {Roskos1992} with quantitative theoretical methods, we consider a numerical setup in which $H$ is given by formula (\ref {F2}), with
\be
\ell_{w_1}= 145 \mbox {\AA}\, , \ \ell_{w_2}=100 \mbox {\AA}\, , \  \ell_{b_1} = 25 \mbox {\AA} \, , \ \mbox { and } \ s=0.2\, .
\ee
When the electric field is turned on, i.e., $F>0$, $H$ no longer possesses eigenvalues, supporting quantum resonances instead; these are solutions to the time-independent equation $H\psi = E\psi$ such that $\Im E <0$ and satisfying the outgoing conditions $\psi (x) \sim 0$, as $x \to - \infty$, and \cite {Siegert1939,Korsch2003}
\bee
 \psi (x) \sim iA_i (\zeta_5 ) + B_i (\zeta_5 )\, \ \mbox { as } \ x\to + \infty \, . \label {F5}
\eee
The general solution to equation $H\psi =E\psi$ is given by
\be
\psi (x)=a_j A_i (\zeta_j ) +b_j B_i (\zeta_j ) \ \mbox { if } \ x \in (x_{j-1} , x_j ) \, ,
\ee
where $j=1,\ldots ,5$, $A_i$ and $B_i$ are the Airy's special functions,
\be
\zeta_1 =\zeta_3 =\zeta_5 :=  -\left ( \frac {2 m_b^\star }{\hbar^2 e^2 F^2} \right )^{1/3} (eF x+E)
\ee
and
\be
\zeta_2 =\zeta_4 :=  -\left ( \frac {2 m_w^\star }{\hbar^2 e^2 F^2} \right )^{1/3} (eF x+E-v)\, . 
\ee
Then we impose the BenDaniel and Duke matching conditions \cite {BenDaniel1966}
%
%\begin{widetext}
\be
\psi (x_j+0) = \psi (x_j-0) \ \mbox { and } \ \frac {1}{m_{j+1} } \frac {\partial \psi (x_j+0)}{\partial x } = \frac {1}{m_{j} } \frac {\partial \psi (x_j-0)}{\partial x } \, , j=1,2,3,4 \, ,
\ee
%\end{widetext}
%
and the outgoing conditions (\ref {F5}), that is
\be
b_1=0 \ \mbox { and }\ a_5 =1 \, , \ b_5=i  \, .
\ee

\begin{figure}
\begin{center}
\includegraphics[height=4.2cm,width=4.2cm]{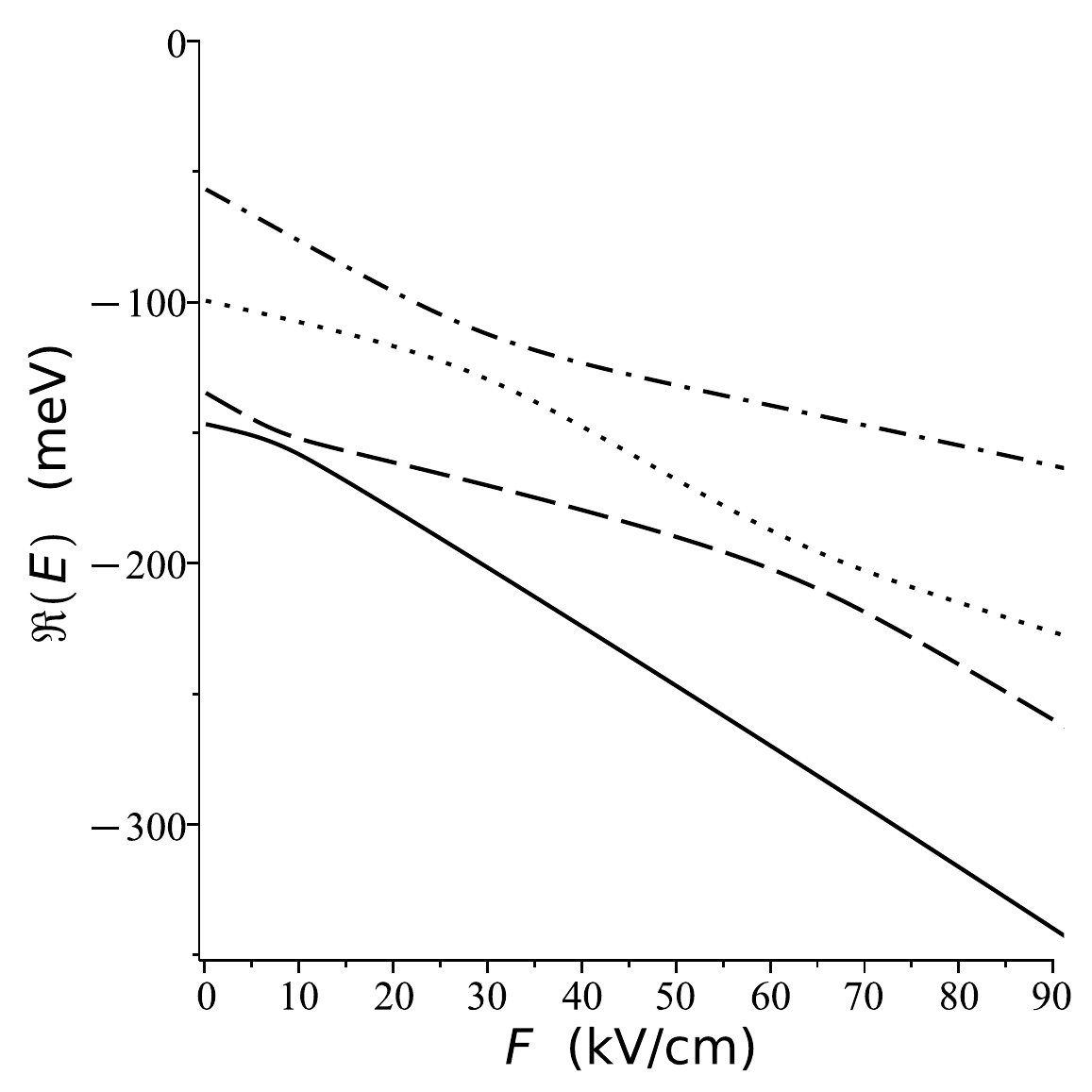}
\includegraphics[height=4.2cm,width=4.2cm]{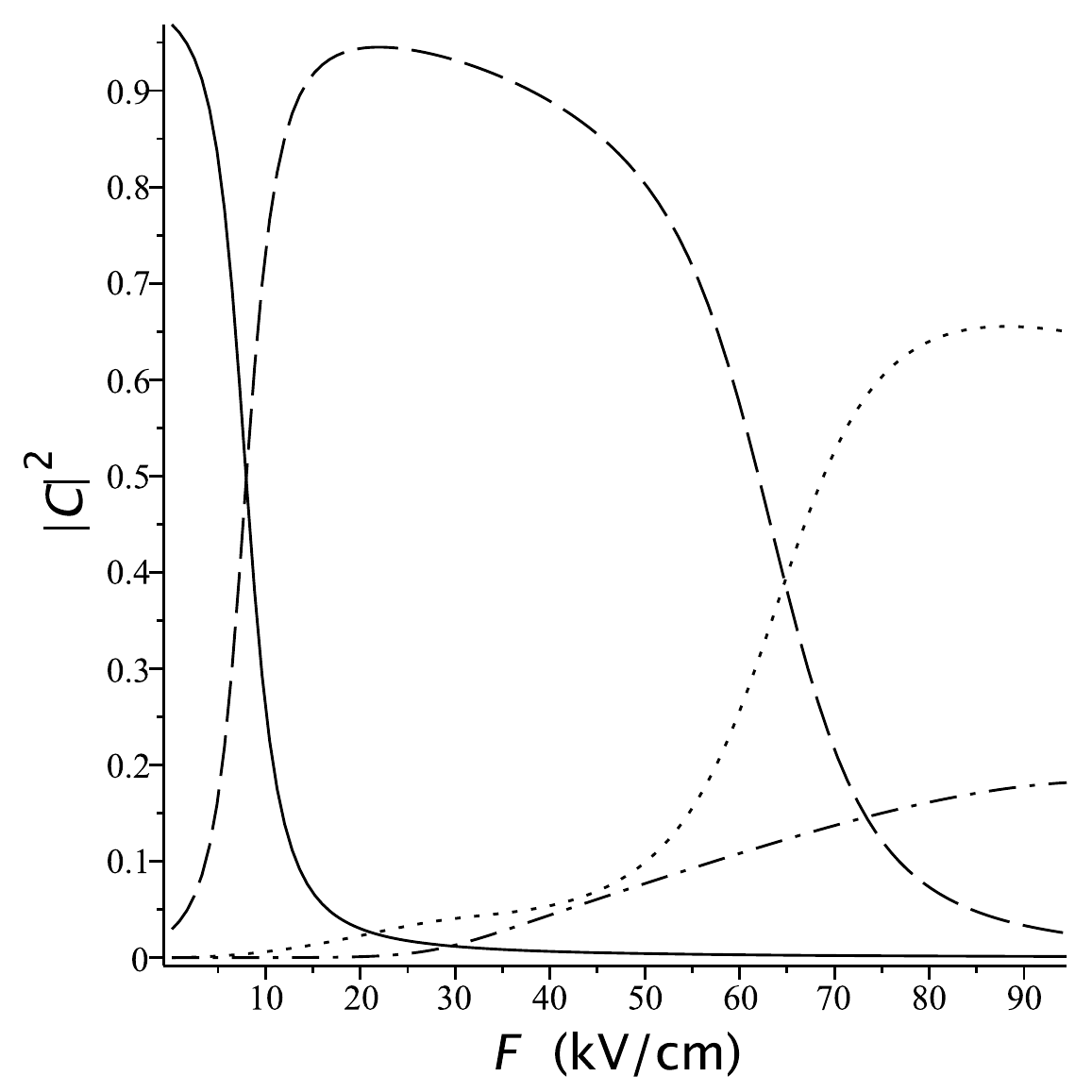}
\caption{Left-hand panel: avoided crossings of the real parts of the four resonances $E_1$ (solid line), $E_2$ (dashed line), $E_3$ (dotted line), and $E_4$ (dash-dotted line), showing three Type~$I$ RCs at $F=7.9(6)$ kV/cm, $F \approx 30$ kV/cm, and $F \approx 65$ kV/cm. Right-hand panel: squared absolute values $|C_{E_j}|^2$ ($j=1,2,3,4$) of the expansion coefficients in formula (\ref {F6}). \ The model parameters are taken from \cite{Roskos1992}.}
\label {Fig2}
\end{center}
\end{figure}

If we denote by $N_{ris}$ the number of quantum resonances with a negative imaginary part sufficiently small in absolute value, then, for the range of values of $F$ considered, we find that $N_{ris}=4$. \ It is worth noting in the left-hand panel of Fig.~\ref{Fig2} that the first two resonances exhibit a narrow Type~$I$ RC.\ That is, their real parts show an avoided crossing at $F=F_C$, where $F_C = 7.9(6)$ kV/cm is the crossing point value; at this point, the imaginary parts of the two quantum resonances intersect exactly, and a beating motion of the wavefunction between the two wells is therefore expected. \ Two further Type~$I$ RCs occur, but they are not narrow: the first is between the third and fourth resonances at the crossing point value $F_C \approx 30$ kV/cm, while the second is between the second and third resonances at the crossing point value $F_C \approx 65$ kV/cm.

Since $H$ is a self-adjoint operator (see Appendix \ref {App2} ) then the dynamics of physical observable, like the \emph {survival amplitude}
\be
\A (t) = \langle \psi_0 | e^{-itH/\hbar } \psi_0 \rangle \, ,
\ee
can be computed (see, e.g., formula (\ref {F23})) as follows
\bee
\A (t) \approx \sum_{j=1}^{N_{ris}} C_{E_j} e^{-itE_j /\hbar } \label {F6}
\eee
where the coefficients $C_{E_j}$ depend on the DC electric field strength as shown in Fig. \ref {Fig2}, right-hand panel.

As already noted, in the case of two resonances that undergo a Type~$I$ RC, when $F$ varies within a small neighborhood of the crossing point $F_C$, the associated metastable states are equally localized in both wells.\ Since the wavefunction is initially prepared in the wider well, it follows from Eq.~(\ref {F26}) that the numerical coefficients $C_{E_j}$ associated with the aforementioned two resonances have almost equal absolute values for $F$ close to $F_C$.\ This fact is evident in the right-hand panel of Fig.~\ref {Fig2}: when $F$ varies over an interval centered at $F_C = 7.9 (6)$ kV/cm, $E_1$ and $E_2$ exhibit a Type~$I$ RC, and the squared absolute values of the associated coefficients $C_{E_1}$ and $C_{E_2}$ are both close to $1/2$.\ In this case, an interference effect is triggered in (\ref {F6}), and large-amplitude oscillations of $\A (t)$ with period $T=2\pi \hbar /( \Re E_2 - \Re E_1 )$ are expected.\ A similar scenario occurs when $F$ varies over an interval centered at $F_C \approx 65$ kV/cm.\ On the other hand, far from a crossing point $F_C$, or in the case of a Type~$II$ RC, the metastable states associated with the resonances are localized within a single well, and, as can be seen in Fig.~\ref {Fig2}, only one of the numerical coefficients $C_{E_j}$ is non-negligible.\ Consequently, no interference effect takes place.\

In conclusion, the theoretical model predicts that large-amplitude periodic oscillations of $\A (t)$ occur when the electric field takes values between $5$ and $15$ kV/cm, with frequencies ranging from $1.25$ THz (when $F=F_C=7.9 (6)$ kV/cm) up to $2$ THz. \ The obtained theoretical results are therefore in full agreement with the experimental observations reported in \cite {Roskos1992}. \ Furthermore, additional oscillations are predicted when the electric field strength is on the order of $60$ kV/cm; however, these findings currently lack an experimental counterpart.

\section {Results: tuning the model parameters
%Proposal of a quantum sensor for the DC electric field
} \label {Sec3}
We can now treat the experiment as an inverse problem by assuming the electric field strength $F$ is unknown, which is precisely the quantity we aim to measure. \ If we observe sufficiently large oscillations, we can then conclude that the applied field falls within a range centered at $F_C = 7.9(6)$~kV/cm or $F_C \approx 65$~kV/cm.\ Hence, the device acts as a sensor to verify whether the applied field is close to one of these critical values $F_C$. \ There are, however, two serious shortcomings if one wants to use this device for this purpose: two distinct ranges of field values give rise to oscillations rather than just one, and, more importantly, these ranges are far too broad (the first, for instance, extends from $5$ to $15$~kV/cm). \ It is therefore necessary to design a device featuring a single critical value $F_C$ and a much narrower oscillation region centered around it. \ These two goals can be achieved by appropriately tuning the widths of the two wells and the internal barrier. \ In particular, by appropriately decreasing the width of the two wells, the number of resonances is reduced to only two; furthermore, by widening the barrier between the wells, the two resonances become increasingly narrow, that is their imaginary parts become very small in absolute value, as does the splitting of their real parts near the crossing point. \ Care must clearly still be taken to ensure that these parameters are adjusted so that the RC remains of Type~$I$ rather than Type~$II$; this can be verified by calculating the behavior of the quantum resonances as the field varies, or by checking that the condition $\rho_i / 2\rho_e < 1$ is satisfied~\cite{Grecchi1995}, where $\rho_i$ and $\rho_e$ are the Agmon metric measures of the internal and external barriers, respectively.

\begin{figure}
\begin{center}
\includegraphics[height=4.2cm,width=4.2cm]{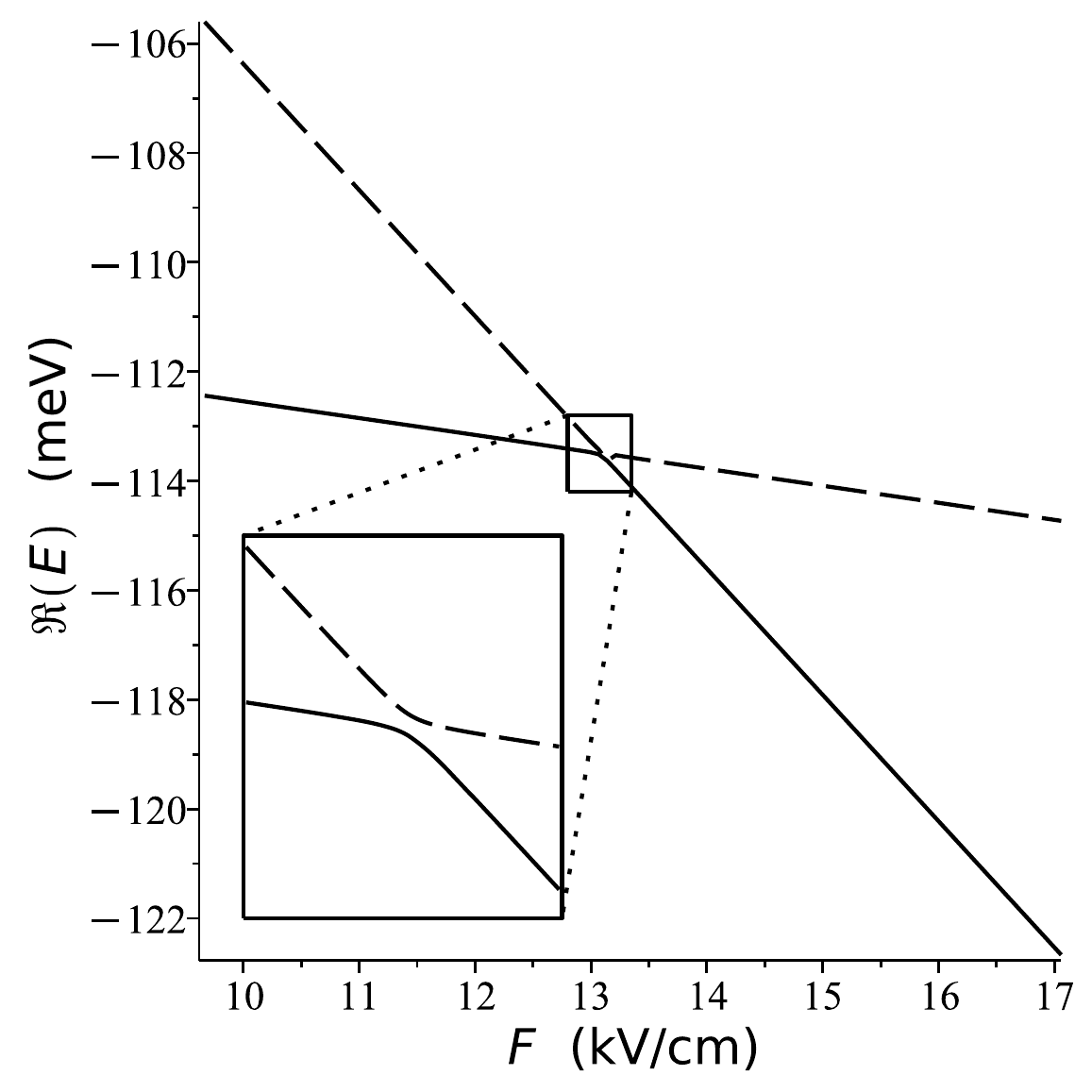}
\includegraphics[height=4.2cm,width=4.2cm]{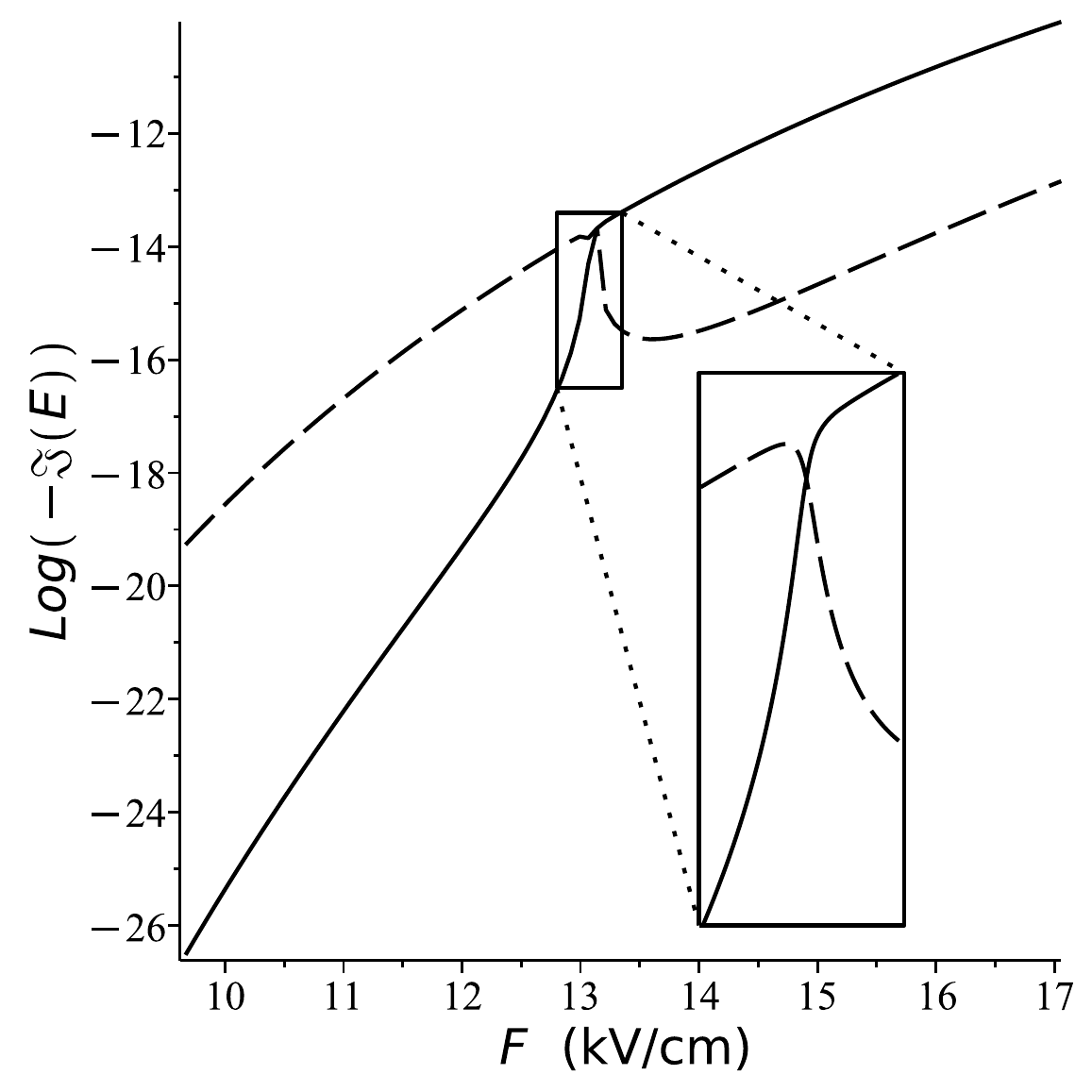}
\caption{Left-hand panel: avoided crossing of the real parts of the two resonances $E_1$ (solid line) and $E_2$ (dashed line). \ Right-hand panel: crossing of the imaginary parts of the two resonances. \ The RC is of Type~$I$ and occurs at $F=F_C =13.087(5)$ kV/cm, with heterostructure parameters given by (\ref {F7}).}
\label {Fig3}
\end{center}
\end{figure}

By making use of the theoretical analysis developed above, we find that these two goals can be achieved, for example, by choosing the heterostructure parameters as follows:
\bee
\ell_{w_1}= 60\, \mbox {\AA} \, , \ \ell_{w_2}=40\, \mbox {\AA} \, , \ \ell_{b_1} = 150\, \mbox {\AA} \ \mbox { and } \ s=0.2\, .\label {F7}
\eee
When the electric field is applied, $H$ has no eigenvalues but, for the range of values of $F$ considered in our numerical experiment, possesses only $N_{ris}=2$ quantum resonances which, as shown in Fig.~\ref {Fig3}, exhibit a narrow Type~$I$ RC at the crossing point $F_C=13.087(5) $ kV/cm.

\begin{figure}
\begin{center}
\includegraphics[height=4.2cm,width=4.2cm]{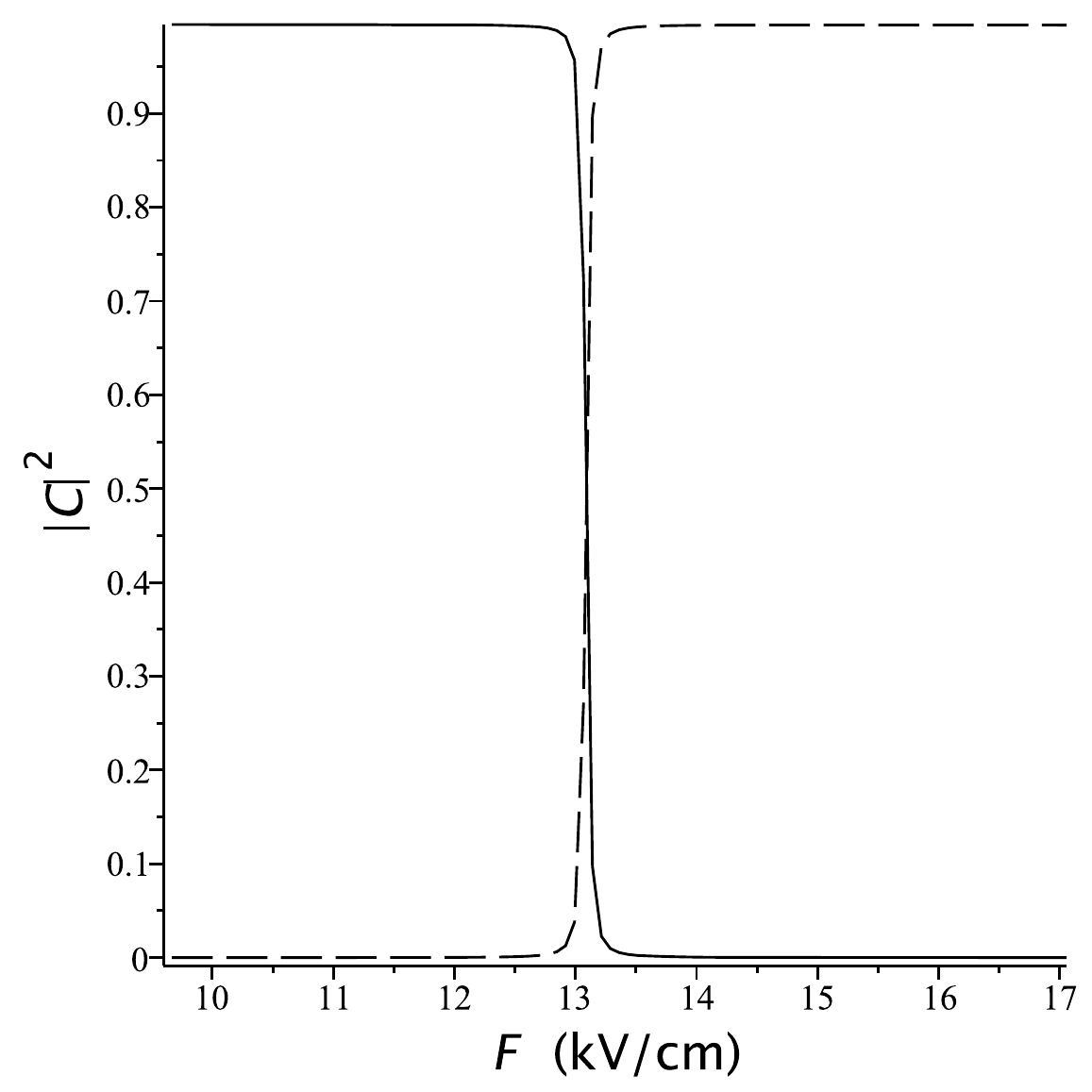}
\includegraphics[height=4.2cm,width=4.2cm]{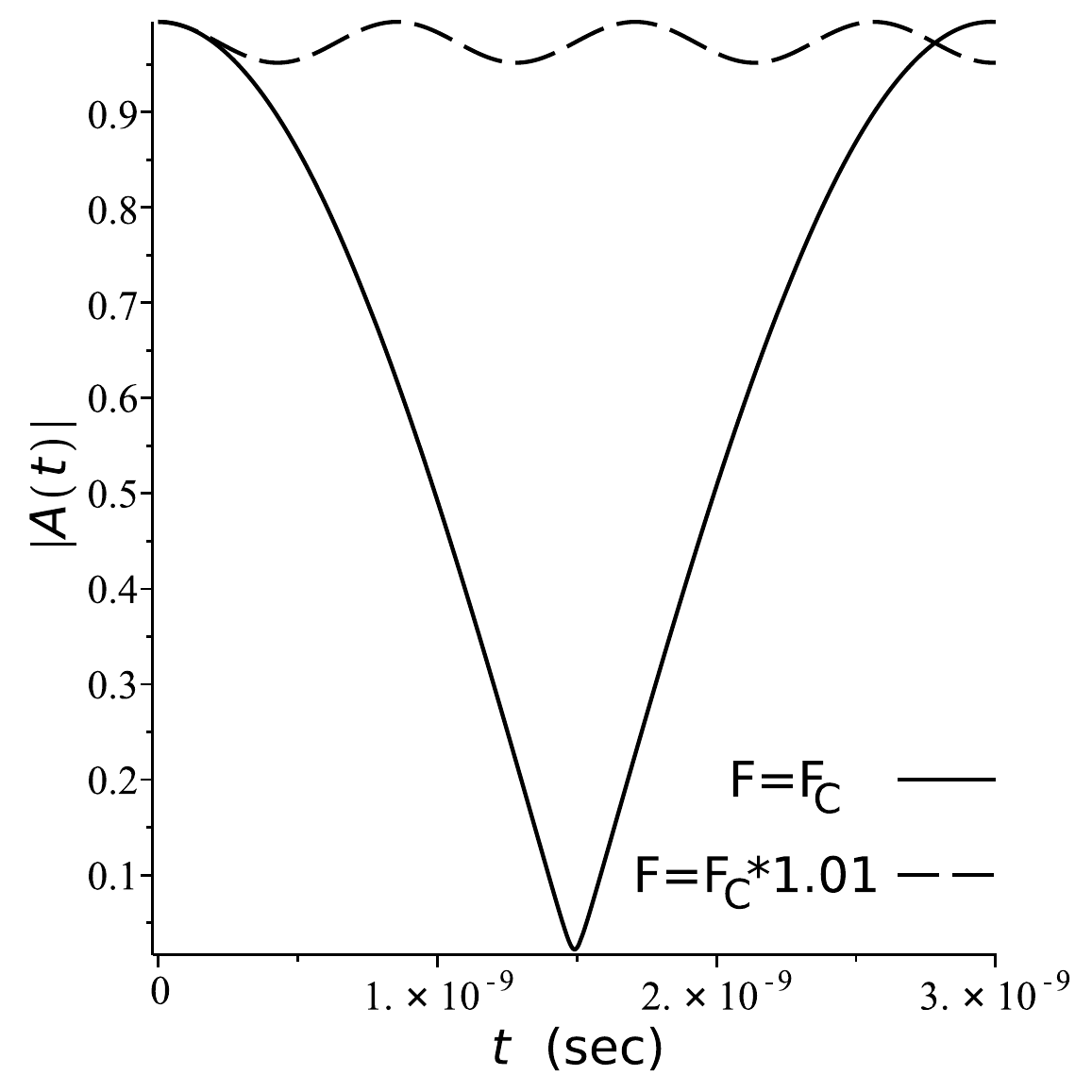}
\caption{The squared absolute values of the coefficients $C_{E_1}$ (solid line) and $C_{E_2}$ (dashed line) (left-hand panel) and the absolute value of $\A(t)$ for an initial wavefunction localized in just one well (right-hand panel), for model parameters given by (\ref {F7}). \ A complete beating motion occurs for $F=F_C$ (solid line), whereas no large-amplitude oscillations appear for $F=F_C \cdot 1.01$ (dashed line).}
\label {Fig4}
\end{center}
\end{figure}

For a wavefunction initially localized in the wider well, and given that the RC is of Type~$I$, we expect, according to formula (\ref {F26}), that both numerical coefficients $C_{E_j}$ have squared absolute values close to $1/2$ (see the left-hand panel of Fig.~\ref {Fig4}) when $F$ varies within a small neighborhood of $F_C$, in agreement with the numerical results collected in the top table of Table~\ref {Tab1}. \ On the other hand, when $F$ is sufficiently far from the crossing point $F_C$, we expect one of the numerical coefficients $C_{E_j}$ to have an absolute value of virtually $0$ (or $1$) and the other to be $1$ (or $0$) (see the left-hand panel of Fig.~\ref {Fig4}); the numerical results collected in the lower table of Table~\ref {Tab1} agree with this picture for $F=F_C \cdot 1.01$.
\begin{table}[h]
\centering
\begin{tabular}{ll}
\hline
 \multicolumn{2}{c}{$F=F_C =  13.087(5) $ kV/ cm}  \\
\hline
$E_1= -113.535(3) - i 0.101(7) \cdot 10^{-13} $
&  $ |C_{E_1}|^2 = 0.486(4)   $    \\
$E_2= -113.456(5) -i  0.101(8) \cdot 10^{-13} $
& $  |C_{E_2}|^2 = 0.508(3)   $  \\
\hline \hline
 \multicolumn{2}{c}{$F=F_C\cdot 1.01$}  \\
\hline
 $E_1=  -113.804(9) - i  0.284(2) \cdot 10^{-13} $
 & $  |C_{E_1}|^2 = 0.0214(5) $ \\
 $E_2 = -113. 529(7)- i 0.732(1) \cdot 10^{-15} $
& $ |C_{E_2}|^2= 0.973(2) $   \\
 \hline
\end{tabular}
\caption{Resonance energies $E_j$ (in meV) and coefficients $C_{E_j}$ ($j=1,2$) for $F=F_C$ and $F=F_C \cdot 1.01$. The model parameter values are given by (\ref {F7}).}
\label {Tab1}
\end{table}

Consequently, the behavior of the survival amplitude $\A(t)$, calculated using (\ref {F6}), is highly sensitive to $F$. \ In fact, for $F=F_C$, the two coefficients $C_{E_1}$ and $C_{E_2}$ are of the same order of magnitude, triggering a significant beating effect with period $T=0.526(8) \cdot 10^{-10} \, \mbox {sec.}$ (see the right-hand panel of Fig.~\ref {Fig4}). \ On the other hand, if $F=F_C \cdot (1 + \epsilon)$, with a relative difference of $\epsilon = 10^{-2}$, no large-amplitude oscillations occur (see the right-hand panel of Fig.~\ref {Fig4}) because the imaginary parts of the two resonances are quite different and, more importantly, one of the two coefficients is virtually zero, while the other one is virtually equal to $1$ in absolute value.
\begin{figure}
\begin{center}
\includegraphics[height=6cm,width=6cm]{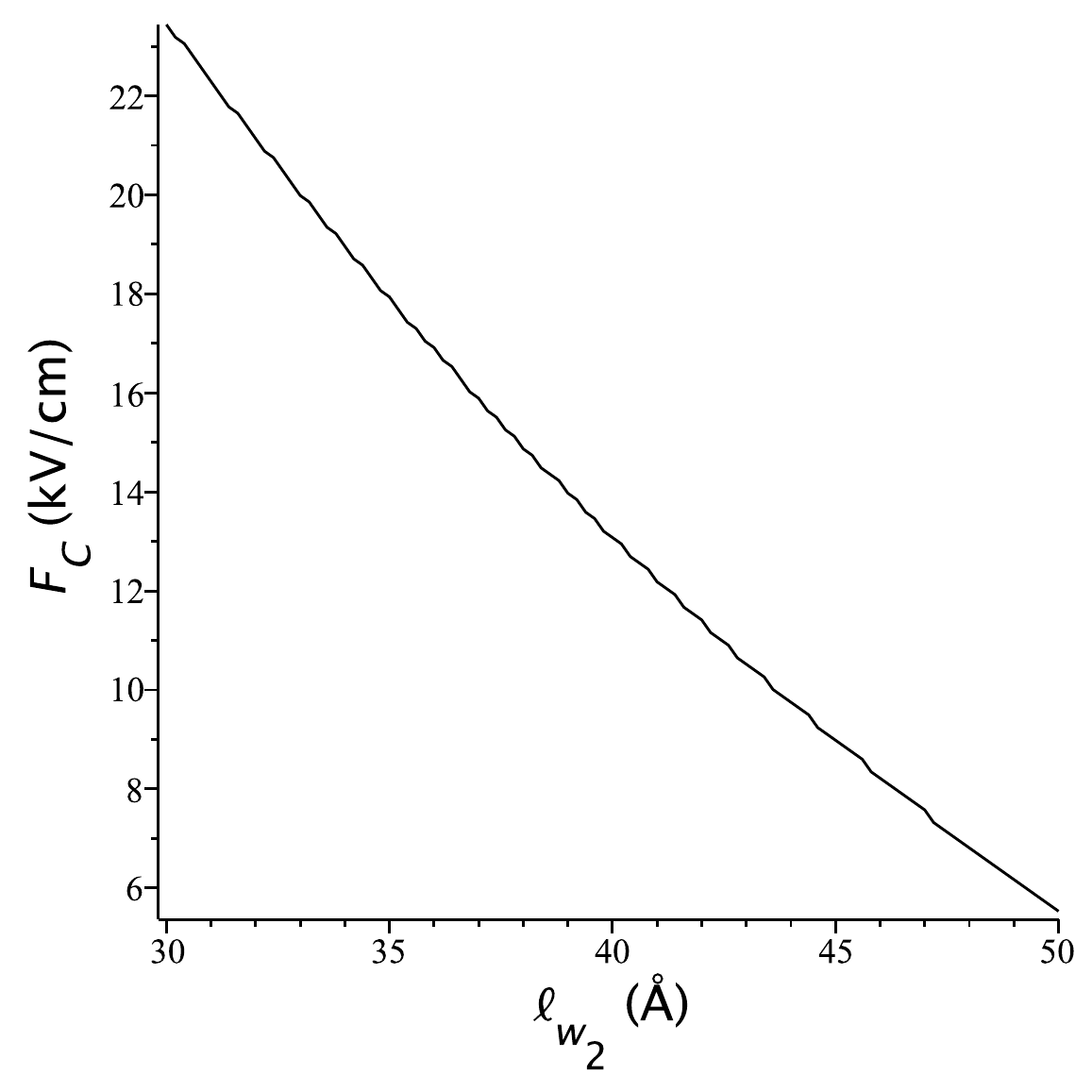}
\caption{The value of the crossing point $F_C$ as function of the width $\ell_{w_2}$. \ The values of the other parameters $\ell_{w_1}$, $\ell_{b_1}$ and $s$ are those given by (\ref {F7}).}
\label {Fig5}
\end{center}
\end{figure}

In conclusion, the effect of an external DC electric field on the dynamics of the survival probability of a double-well heterostructure for model parameters given by (\ref {F7}) generally consists of producing a small-amplitude oscillation that decays exponentially. \ However, when the electric field strength reaches a critical value corresponding to the crossing point $F_C$, the situation changes rapidly and abruptly; indeed, at this value, the amplitude of the oscillations becomes significant. \ Therefore, this property can be exploited to verify whether the external electric field has a predetermined intensity, with a relative error that is of the order of $1\%$. \ It is worth noting in Fig.~\ref{Fig5} that the critical value $F_C$ can be selected in the range between $6$~kV/cm to $23$~kV/cm by appropriately adjusting the widths of the two wells and the internal barrier (taking care to maintain a Type~$I$ RC). \ We also point out that it is possible to conceive  devices with tighter tolerances; for example, it can be verified that for $s=0.3$ (again with $\ell_{w_1}= 60\, \mbox {\AA}$, $\ell_{w_2}=40\, \mbox {\AA}$ and $\ell_{b_1} = 150\, \mbox {\AA}$), the crossing point occurs at $F=F_C=17.285(2)$ kV/cm and that the resulting device has a relative error of the order of $0.05\%$.

\section{Conclusion} \label {Sec4}
Quantum sensors are devices that measure physical quantities, such as electric or magnetic fields, by exploiting the laws of quantum mechanics. \ For instance, a Ramsey interferometer fundamentally consists of a two-level quantum system perturbed by the external field to be measured \cite{Ramsey,Degen}. \ Let $\omega$ be the transition frequency between two unperturbed energy levels, typically two ground-state levels; an interference effect then generates periodic oscillations with period $T = 2\pi \hbar / \omega$. \ The key idea is that the perturbation induces a shift in the energy levels, leading to a variation of the oscillation period. \ Consequently, the relationship between the oscillation period and the external field allows one to estimate the field strength: by measuring the period with sufficient precision, the intensity of the external field can be inferred. \ Currently, many other quantum sensors for DC electric fields have been developed, from those based on silicon carbide to those based on nitrogen-vacancy centers in nanodiamonds  (see \cite {Dolde2011,Gaborit2013,Hollendonner2023,Castelletto2024,Katsumi2025} and references therein).

Based on the previous analysis, a theoretical model is conceived for a quantum sensor utilizing asymmetric double-well heterostructures perturbed by an external DC electric field with a strength on the order of some~$\text{kV/cm}$. \ The energy levels corresponding to the ground states become quantum resonances and exhibit a resonant crossing (RC) when the external electric field strength reaches a predefined value. \ If the RC is of Type~$I$, a periodic spatial oscillation of the wavefunction, initially localized in the wider well, occurs between the two wells at the crossing point. \ Consequently, one can easily determine whether the strength of an external DC electric field is near this predefined target value simply by detecting the presence or absence of these oscillations, without needing to measure their period or perform any calculations. \ The primary advantages of such a device, compared to both conventional and recently developed sensors, would be its operational simplicity and rapid response time.

Our proposal  is, in a sense,  a quantum counterpart to the classical resonance effect observed in driven RLC circuits, where matching the driving frequency to an intrinsic frequency induces large-amplitude oscillations. \ A similar mechanism governs the quantum heterostructure described above. \ Provided that a Type~$I$ RC takes place when $F$ matches a predefined value $F_C$, where the energy levels are nearly aligned, resonant tunneling causes a wavepacket initially prepared in one well to oscillate between the wells, leading to large-amplitude oscillations in the survival amplitude $\mathcal{A}(t)$. \ Consequently, the device operates as a quantum sensor, yielding an output of "1" when these oscillations occur and "0" otherwise. \ It is worth emphasizing that a generic resonances crossing is not sufficient to observe such large-amplitude beating dynamics; indeed, the crossing must be of Type~$I$ rather than Type~$II$, with the transition between the two regimes passing through diabolical points.

Furthermore, we have demonstrated how to tune the heterostructure parameters to select the target value $F_C$ to be detected, as well as how to estimate the operating tolerance of the device.

We close by emphasizing that this type of device may find several practical applications where the strength of the DC electric field is on the order of some~$\text{kV/cm}$.  \ To name a few, it would allow one to rapidly and straightforwardly determine whether the charge of a nanobattery~\cite{Kim2004,Vullum2006,Porz2017,Ding2024} drops below (or exceeds, during charging) a predetermined threshold. \ In medical diagnostics, it could be applied to ultra-dense electric field encephalography~\cite{Petrov2014} to monitor electrical activity in specific brain regions, or to enable the detection of electric fields on biological nanosurfaces~\cite{Sel2005,Liu2012,Aslam2023}. \ Other potential application areas include plasma physics, with particular emphasis on plasma--liquid interactions~\cite{Kovacevic2022}; polymer technologies, where electric fields are used to create charged jets~\cite{Doshi1995}; and, finally, medical treatments and therapies based on electroporation~\cite{Miklavcic1998,Davalos2005,Corovic2013}.

\thanks {Andrea Sacchetti acknowledges the Gruppo Nazionale per la Fisica Matematica of Istituto Nazionale di Alta Matematica (GNFM-INdAM). \ This work was partially supported by the Next Generation EU - Prin 2022CHELC7 project "Singular Interactions and Effective Models in Mathematical Physics"  and by the UniMoRe-FIM project  "Modelli e metodi della Fisica Matematica''.}

%\printbibliography[heading=subbibliography, title={Main References}]

%\end{refsection}

%\begin{refsection}

\appendix

\section {Effective mass approximation} \label {App1}
Here, to introduce the envelope function approximation, we briefly summarize a few key results; we refer to \cite {Altarelli1986}, see also \cite {Bastard1990,Harrison2005}, for details.

\subsection {Crystal momentum representation} Let us consider the time-dependent Schr\"odinger equation for an electron with mass $m_0$ in a periodic crystal with potential $V_{per} (\vvr )$, $\vvr \in \R^3$, and an external potential $V(\vvr )$:
\be
i \hbar \frac {\partial \psi}{\partial t} = H \psi \, ,\ \psi = \psi (\vvr , t) \, , \ \vvr \in \R^3 \, , \ t \in \R \, ,
\ee
where $H= H_B + V(\vvr ) $ and $H_B = - \frac {\hbar^2}{2m_0} \Delta_{\vvr}  + V_{per}( \vvr ) $. \
The crystal potential $V_{per}$ is a periodic function where $\vR$ is the lattice vector: $V_{per}(\vvr ) = V_{per}(\vvr + \vR )$, $\forall \vvr \in \R^d$. \ It is well known that the spectrum of the Bloch operator $H_B$ is given by a sequence of bands: $\sigma (H_B) = \cup_{n=1}^\infty [\E_n^b, \E_n^t] $ where
\be
\E_n^b = \min_{\vk \in \B} \E_n (\vk ) \ \mbox { and } \ \E_n^t = \max _{\vk \in \B} \E_n (\vk ) \, ,
\ee
where $\B$ is the Brillouin zone and $\vk$ is the quasi-momentum variable. \ The band functions $\E_n (\vk )$ are periodic functions $\E_n (\vk + \vG )=\E_n (\vk )$ where $\vG$ is a vector of the reciprocal lattice. \ We denote by  $\psi_n (\vk , \vvr )$ the Bloch functions such that $ H_B \psi_n (\vk , \vvr )= \E_n (\vk ) \psi_n (\vk, \vvr ) $ and $ \psi_n (\vk , \vvr )= e^{i \vk \cdot \vvr } u_n (\vk , \vvr ) $ where $u_n (\vk , \vvr ) = u_n (\vk , \vvr + \vR)$ is a periodic function with the same period of the crystal lattice. \ By construction $\psi_n (\vk + \vG ,\vvr ) = \psi_n (\vk , \vvr )$.

The sequence of Bloch functions is a complete orthonormal set of functions and any wavefunction $\psi (\vvr , t) $ can be represented as follows
\be
\psi (\vvr , t) = \sum_{n=1}^\infty \int_{\B} \phi_n (\vk , t) \psi_n (\vk , \vvr ) \, d \vk \, ,
\ee
where
\be
\phi_n (\vk  ,t) = \langle \psi_n (\vk , \cdot )  | \psi (\cdot , t ) \rangle  \, .
\ee
The set of functions $\{ \phi_n (\vk , t) \}_{n=1}^\infty $ is denoted the Crystal Momentum Representation (CMR) of the wavefunction $\psi (\vvr , t) $. \

It is easy to check that in the CMR the time-independent Schr\"odinger equation $(H-E)\psi =0$  takes the form
\bee
\E_n (\vk ) \phi_n (\vk ) + \sum_{m=1}^\infty \int_{\B} \langle \psi_n (\vk , \cdot) , V (\cdot ) \psi_m (\vk ', \cdot ) \rangle \phi_m (\vk ')\, d \vk '= E  \phi_n (\vk ) \, , \ n\in \N  \, ,\label {F12}
\eee
and similarly the time-dependent one becomes
\bee
i \hbar \frac {\partial \phi_n (\vk , t)}{\partial t} = \E_n (\vk ) \phi_n (\vk ,t) + \sum_{m=1}^\infty \int_{\B} \langle \psi_n (\vk , \cdot) , V (\cdot ) \psi_m (\vk ', \cdot ) \rangle \phi_m (\vk ',t) \, d \vk ' \, , \ n\in \N \, .\label {F13}
\eee

Physical observable can be represented by means of the CMR. \  In particular, let us consider the survival amplitude
\be
\A (t) = \langle \psi (\cdot , 0) | \psi (\cdot , t ) \rangle
:= \int_{\R^3} \overline { \psi (\vvr , 0) } \, \psi (\vvr , t ) \, d \vvr
\, ,
\ee
where $\psi (\vvr , 0)$ is the quantum wavefunction at the initial time $t_0=0$, $\psi (\vvr ,t)$ is the quantum wavefunction at the instant $t$. \ By means of a straightforward calculation and recalling the orthonormal property then it follows that the survival amplitude can be computed by means of the CMR as follows
\be
\A (t) = \sum_{n=1}^\infty \int_{\B } \overline {\phi_n (\vk , 0 ) } \phi_n (\vk ,t) \, d \vk \, .
\ee

\subsection {Effective mass approximation in the bulk}
Some assumptions are required. \ The first one concerns the spectrum of the Bloch operator. \ Indeed, we assume that one band, say the first one, is separated from the remainder spectrum; that is the width of the first gap $\E_2^b - \E_1^t$ is assumed to be strictly positive and large enough.

The second one concerns the external potential. \ That is, we assume that the external potential $V (\vvr )$ is a slowly varying function over the periodical crystal lattice.

Under these two assumptions it follows that in the CMR the restriction of the time-independent Schr\"odinger equation to the first band leads to the following single equation
\bee
\E_1 (\vk ) \phi_1 (\vk ) + \int_{\B} \tilde V (\vk -\vk ') \phi_1 (\vk ') \, d \vk '= E  \phi_1 (\vk )\label {F14}
\eee
where $\tilde V$ is the Fourier transform of  $V$. \ In the case of an isotropic non-degenerate band extremum we  make use of the quadratic approximation
\be
\E_1 (\vk ) \approx \E_1 (0) + \frac {\hbar^2}{2m^\star} |\vk |^2\, ,
\ee
where $m^\star$ is the effective mass, and,
then equation (\ref {F14}) becomes
\be
\E_1 (0) \phi_1 (\vk )  + \frac {\hbar^2}{2m^\star} |\vk |^2 \phi_1 (\vk ) + \int_{\B} \tilde V (\vk -\vk ') \phi_1 (\vk ') d \vk '= E  \phi_1 (\vk )\, .
\ee
by applying the Fourier transform to both sides of (\ref {F14})
 then this equation takes the final form
\bee
- \nabla \left [ \frac {\hbar^2}{2m^\star} \nabla \right ] \Phi_1 + [ \E_1 (0) + V (\vvr )] \Phi_1   = E  \Phi_1 \, , \label {F15}
\eee
usually called the time-independent \emph {effective mass equation}, where   $\Phi_1 (\vvr )$ denotes the Fourier transform of $\phi_1 (\vk )$. \ Similarly,  the time-dependent equation effective mass equation follows (for details see \S 7.6 \cite {Jacoboni2010}, see also \cite {Muga1992,Cruz1993}).
\bee
i \hbar \frac {\partial  \Phi_1 }{\partial t}  =  -  \nabla \left [ \frac {\hbar^2 }{2m^\star} \nabla \right ]  \Phi_1  +  [\E_1 (0) + V (\vvr )] \Phi_1   \, .  \label {F16}
\eee

We conclude by representing the wavefunction $\psi (\vvr ,t)$ by means of the solution $\Phi_1 (\vvr ,t)$ to the effective mass equation (\ref {F16}); indeed, by neglecting the other bands, but the first one, and by means of the first order approximation $u_1 (\vk , \vvr ) \approx u_1 (0,\vvr )$, we get that
\bee
\psi (\vvr ,t)\approx   \Phi_1 (\vvr ,t) u_1 (0, \vvr) \, . \label {F17}
\eee

Hence, the survival amplitude takes the form
\be
\A (t)  \approx
\int_{\R^3} \overline {\Phi_1 (\vvr ,0)} \Phi_1 (\vvr , t) |u_1 (0, \vvr )|^2 \, d \vvr \, .
\ee
Since $\Phi_1 (\vvr )$ is a slowly varying \emph {envelope function} modulating the rapidly varying Bloch part $u_1 (0, \vvr )$ then
\bee
\A (t) \approx 
C \int_{\R^3} \overline {\Phi_1 (\vvr ,0)} \Phi_1 (\vvr , t)  \, d \vvr \label {F18}
\eee
for some positive constant $C = \int_{\Lambda} |u_1 (0, \vvr )|^2 \, d \vvr$ where $\Lambda$ is the bulk unit cell. \ We always can assume that $C=1$ by means of a suitable normalization of $\Phi_1 (\vvr , 0)$.

\subsection {Effective mass equation for heterostructures}

In the case of heterostructures we have planar layers of two, or more, different materials (typically semiconductors) connected to each other by heterojunctions. \ Assume now, for argument's sake, that the direction along which the different layers of material change is the $x$ direction, where $\vvr =(x,y,z)$. \ Therefore, in effective mass equation (\ref {F15}) and (\ref {F16}), both the effective mass $m^\star$ and the bottom energy value $\E_1 (\vk )$ at $\vk =0$ will depend on $x$ and they will be denoted $m^\star (x)$ and $\E_1 (0,x)$.

We underline that the functions $m^\star (x)$ and $\E_1 (0,x)$ have a discontinuity at the heterojunction planes. \ In fact, in order to preserve the continuity of the density of the current, the continuity property of the following functions
\be
\Phi_1 (\vvr ) \ \mbox { and }
\frac {1}{m^\star (x)} \frac {\partial  \Phi_1 (\vvr )}{\partial x}
\ee
must hold true at the heterojunction planes \cite {BenDaniel1966}. \ In particular, let us now consider an heterojunction between two different types, named $A$ and $B$, of semiconductors that interface at the plane $x=0$. \ We denote by $\E_1^{X}(\vk )$ and by $\psi_1^{X} (\vk ,\vvr) =e^{i \vk \cdot \vvr} u_1^{X} (\vk , \vvr )$ the first band function and the first Bloch function of the semiconductor $X=A,B$. \ Therefore, in the half-space $x<0$, equations (\ref {F15}) and (\ref {F16}) will apply with a certain effective mass $m_A^\star$ and a certain bottom value $\E_1^A (0) $; while in the half-space $x>0$, they will apply with different effective mass $m_B^\star$ and value $\E_1^B (0)$. \ To connect the two equations, we should also assume that $u_1^A(0) \approx u_1^B (0)$; we should emphasize that this condition holds true for the family III-V of semiconductors as long as we consider the conduction band minimum at both sides \cite {Altarelli1986}. \ Then the matching conditions at the interface, to ensure the continuity of the current density, are the continuity of $\Phi_1 (\vvr )$ and the so-called BenDaniel and Duke conditions \cite {BenDaniel1966}
\bee
\lim_{\epsilon \to 0^+} 
\frac {1}{m^\star_A} \frac {\partial  \Phi_1 ( -\epsilon ,y,z)}{\partial x} = \lim_{\epsilon \to 0^+} 
\frac {1}{m^\star_B} \frac {\partial  \Phi_1 (\epsilon ,y,z )}{\partial x} 
\, . \label {F19}
\eee

\section {Self-adjoint extension of the Schr\"odinger operator with non-constant mass} \label {App2}

Here, we consider mathematical results concerning time-dependent Schr\"o\-din\-ger equations of the type (\ref {F16}) in any dimension $d=1,2,\ldots $
\bee
\left \{ 
\begin {array}{l} 
i \hbar \frac {\partial \psi}{\partial t} = H \psi \\
\left. \psi (\vvr , t ) \right |_{t=0} = \psi_0 (\vvr )
\end {array}
\right. 
\, , \ \psi (\cdot , t) \in L^2 (\R^d )  \, , \label {F20}
\eee
where $\psi_0 (\vvr ) = \psi (\vvr , 0)$ is the initial wavefunction associated to an Hamiltonian function, $H=T+V$, where
\bee
T = - \nabla_\vvr \left [ p(\vvr ) \nabla_\vvr \right ] \, \ p(\vvr ) =  \frac {\hbar^2}{2m(\vvr )} \, , \label {F21}
\eee
is the kinetic operator, the mass $m (\vvr )$ depends on the spatial variable $\vvr$ and $V(\vvr )$ is a generic potential energy function. In the particular case of equation  (\ref {F16}) actually $m(\vvr )=m^\star (x)$ depends only by $x \in \R$ and $V(\vvr )$ is given by $V(\vvr )+ \E_1 (0,x)$. \ In the following, concerning the function $m(\vvr )$, we assume that the function $m(\vvr )$ is a real-valued function such that
\be
c \le m(\vvr ) \le C \, , \ \forall \vvr \in \R^d \, ,
\ee
for some two positive constants $0 < c \le C < +\infty$.

We will prove that the operator formally defined by $H$ admits a unique self-adjoint extension; this result is rather important because from this fact it will follow that the solution to (\ref {F20}) is given by $e^{-itH/\hbar } \psi_0$.

\subsection {Preliminary results and comments}

Problem (\ref {F20}) has been already considered from several authors. \ For instance, \cite{Ich87} proved a well-posedness result in $L^2(\R^d )$ of the Cauchy problem (\ref {F20}). \ In the free one-dimensional case, i.e. $d=1$ and $V(\vvr ) \equiv 0$, \cite {Kane1957} considered the case where the function $m (\vvr )$ is a piecewise constant function; under these assumptions dispersion and Strichartz estimates for the solution $\psi $ to (\ref {F20}) have been proved. \ A similar result has been proved by \cite {Beli2016} in the case where $m (\vvr  )$ is BV-function.

Bearing in mind the physical motivations, in this Section we also assume that $m(\vvr )$ is a piecewise smooth function. \ That is, let $\Gamma = \cup_{i=1}^n \Gamma_i $, where $n \in \N$ and $\Gamma_i \subset \R^d$, $i=1,2,\ldots, n$, are smooth connected and disjoint surfaces in $\R^d$; then we assume that
\be
p (\vvr ) \in C^1 (\R^d \setminus \Gamma )
\ee
and $p (\vvr )$ and its gradient $\nabla_\vvr p (\vvr ) $ may have jump discontinuities across $\Gamma $, with finite one-sided traces.

We should point out that if $m(\vvr ) \in C^1 (\R^d )$ then things become particularly easy; indeed, in this case by making use of the Kato-Rellich theorem \cite {Teschl2014} is trivial to prove that the operator $T$ formally defined by (\ref {F21}) has a self-adjoint extension with domain $H^2 (\R^d )$.

On the other side, if $m (\vvr )$ is a discontinuous function with a jump across the two faces $\Gamma_\pm$ of $\Gamma$ let $\dot T$ be the operator formally defined by (\ref {F21}) on the domain
%
%\begin{widetext}
\be
\D (\dot T) = 
\left \{ 
\psi \in H^1 (\R ) \cap H^2(\R \setminus \Gamma ) \ : \ 
\left. p(\vvr ) \frac {\partial \psi }{\partial n} \right |_{\Gamma_-} =
\left. p(\vvr ) \frac {\partial \psi }{\partial n} \right |_{\Gamma_+}
\right \} 
\ee
%\end{widetext}
%
where $ \frac {\partial }{\partial n} $ is the normal derivative on $\Gamma$. \ Then $\dot T$ is a positive and symmetric operator densely defined on $L^2 (\R^d)$. \ By the Friedrichs extension Theorem a self-adjoint extension, denoted by $T$, there exists with self-adjointness domain such that 
\be 
\D (\dot T) \subseteq \D (T) \subseteq \overline {\D (\dot T)}^{\| \cdot \|_{\dot T}}
\ee
where ${\| \cdot \|_{\dot T}}$ is the norm defined as (hereafter $\langle \cdot , \cdot \rangle$ denotes the usual scalar product in the Hilbert space $L^2 $)
\be
\langle \varphi , \psi \rangle_{\dot T} = \langle \varphi , (\dot T +1 ) \psi \rangle \ \mbox { for } \ \psi, \varphi \in \D (\dot T) \, .
\ee

If we introduce then a real-valued bounded potential $V(\vvr )$ let $\dot H$ be the operator formally defined by $\dot T+V$; then it admits an unique self-adjoint extension $H$ on the domain $\D (H) = \D (T)$.

In fact, in many interesting applications of heterostructures the potential $V(\vvr )$ is not bounded from below, like the case of Stark-type potentials considered in this paper, and thus the above results don't apply and a specific theory must be developed under some assumptions. \ In particular we restrict ourselves to the one-dimensional case, i.e. $d=1$.

\subsection {One-dimensional Stark-type problem} Let us consider now the case of $d=1$ where $x \in \R$ is the spatial one-dimensional variable. \  In this case $\Gamma = \{ a_1,a_2,\ldots , a_n \}$ for some $n \ge 1$. \ Concerning the potential we assume that
\be
V(x) = - eF x + W(x)
\ee
where $e,F \not= 0$ and $W (x)$ is a piecewise continuous and bounded real-valued function with compact support; in our model (\ref {F2}) $W(x)$ is the double square wells potential. \ Hereafter, for argument's sake, we assume that
\be
 e=1 \ \mbox { and } \ F>0 \, . 
\ee
We point out that $V(x)$ is not bounded from below and thus the previous results don't apply. 

The following result holds true.

\begin {theorem} \label {Thm1}
The operator 
\be
H\psi = - \left ( p \psi ' \right )'  - F x \psi + W \psi \, ,\ \mbox { where }\ ' = \frac {d}{dx}\, ,  
\ee
formally defined on $\psi \in C_0^\infty (\R )$ admits a unique self-adjoint extension on the domain 
\be
\D (H)= \left \{ \psi \in L^2 (\R ): \psi , p \psi' \in AC_{loc} (\R ) \wedge  H\psi \in L^2 (\R ) \right \}  .
%\label {F22}
\ee
\end {theorem}

%{\it Proof.} \
The proof is quite standard in the framework of the Sturm-Liouville theory and it follows the lines by \cite {Zettl2005} where $n=1$ and $a_1 =0$.

If $\psi,\, p\psi' \in AC_{loc} (\R )$ then $H$ is defined a.e. and we introduce the maximal domain
\be
\D_{max} = \left \{ \psi \in L^2 \ : \ \psi , p\psi' \in AC_{loc} \ \wedge \ H\psi \in L^2 \right \} \, .
\ee
The \emph {maximal} operator is defined as $H_{max} \psi = H \psi $ on ${\D}_{max}  \equiv \D (H)$. \ In order to define the \emph {minimal} one let
\be
\dot H_{min}\psi = H\psi \, , \ \psi \in \D_{max} \cap C_0 (\R)\, .
\ee
It is possible to prove (see Lemma 10.3.1 by \cite {Zettl2005}) that $\dot H_{min}$ admits a symmetric closure; we define the minimal operator $H_{min}$ as the closure of $\dot H_{min}$. \ In fact, by construction it follows that
\be
H_{min} \subseteq H_{max} \, ,
\ee
and, concerning the adjoint operators it is possible to check that
\be
H_{min}^\star = H_{max} \ \mbox { and } \ H_{max}^\star = H_{min} \, .
\ee
Therefore, if $H$ denotes a self-adjoint extension of $H_{min}$ then
\be
H_{min} \subseteq H = H^\star \subseteq H_{max} \, .
\ee
We compute now the defect indices $d^\pm$ of the minimal operator. \ Let $R_\lambda = \mbox {Range} (H_{min} - \lambda \I )$ and let $ N_\lambda =R_\lambda^\perp$ be its orthogonal space, where $ \lambda \in \C $. \ The defect indices are $d^\pm = \mbox {dim } N_{\pm i}$. \ Since $p$ and $V$ are real-valued functions then $d^+=d^-$ and we simply denote by $d$ their common value. \ If we prove that $d=0$ then (see Theorem 10.4.1 by \cite {Zettl2005}) $H_{min}$ is a self-adjoint operator and it has no other proper self-adjoint extension in $L^2 (\R )$; hence, in this case
\be
H_{min} \equiv H \equiv H_{max}
\ee
and thus the Theorem is proved. \ In order to check that $d=0$ we simply observe that (see Theorem 7.4.3 by \cite {Zettl2005}) the endpoints $\pm \infty$ are Limit Points both. \ $\Box$

\begin {remark}
We underline that any function $\psi \in \D (H)$ is such that $p\psi' $ is a continuous function, hence condition (\ref {F19}) holds true at the points $a_i$, $i=1,2,\ldots , n$.
\end {remark}

\begin {remark}
We underline that the effective mass equation (\ref {F2}) fits the assumptions of Theorem \ref {Thm1} and that the mathematical conditions $\psi , p \psi' \in AC_{loc} (\R )$, introduced in order to preserve the symmetry of the linear operator, are equivalent to the physical BenDaniel and Dude conditions (\ref {F19}), introduced in order to preserve the continuity of the current density.
\end {remark}

\section {Quantum resonances and exterior complex scaling} \label {App3} Since the operator $H$ defined in Theorem \ref {Thm1} is a self-adjoint operator, the solution $\psi (x,t)$ to the Cauchy problem (\ref {F20}) is given by the evolution operator
\be
\psi (x,t) = \left [ e^{-i Ht/\hbar } \psi_0 \right ] (x)
\ee
where $\psi_0$ is the initial wavefunction. \ In particular, since the spectrum of $H$ coincides with the entire real axis, i.e. $\sigma (H) = \R$, the survival amplitude $\A (t)$ can be calculated using quantum resonances, extending the result \cite {Herbst1977} and employing the exterior complex scaling method. \ To this end, we note that in our model (\ref {F2}) the effective mass $m(x)$ is a constant function outside of a compact interval  $I_1=[r_2,r_3]$, where $r_2 < r_3$ are chosen such that  the support of the square double-well potential is contained within $I_1$. \ Let $I_2 =[r_1 , r_4] \supset I_1$ for some $r_1 <r_2 <r_3 < r_4$, let $\eta (x) \in C^2 (\R )$ be an exterior cut-off function such that
%
%\begin{widetext}
\be
0 \le \eta (x) \le 1 \, , \ \forall x \in \R \, , \ \eta (x)=0  \, , \ \forall x \in I_1  \, , \  \mbox { and } \ \eta (x) =1 \, ,  \ \forall x \notin I_2 \, .
\ee
%\end{widetext}
%
Let $\U^\lambda$ be the transformation 
\be
\left [ \U^{\lambda }f \right ] (x) = f \left [x+\lambda \eta (x) \right ]\, ,\ \lambda \in \C \, ;
\ee
it is unitary if $\Im \lambda =0$ with $\Re \lambda$, in absolute value, small enough (hereafter we assume to be in such a condition), then 
\be
H^\lambda :={\U^\lambda}  H \left [ {\U^\lambda}\right ]^{-1} 
\ee
is a non-symmetric operator if $\Im \lambda \not= 0$. \ The essential spectrum of $H^\lambda$ is given by  $\sigma_{ess}(H^\lambda ) = \R -i F \Im \lambda $ and the eigenvalues $E^\lambda$ of $H^\lambda$ such that $\Im E^\lambda \in (- F \Im \lambda ,0)$, for some $\lambda$ such that $\Im \lambda >0$, are defined as the quantum resonances of $H$; they are independent of $\lambda$, as well as $-F \Im \lambda <\Im E<0$, and thus we simply denote them as $E$. \ The associated eigenvectors, that is the solution $w^\lambda \in L^2 (\R )$ to the equation $(H^\lambda - E) w^\lambda =0$, are denoted as metastable states; in fact, they actually depend on $\lambda$ and they are given by $\w^\lambda = \U^\lambda \w $
where $\w \notin L^2 (\R )$ is the solution to the equation $(H-E)\w =0$ with the outgoing Siegert's conditions $\w(x) \sim 0$, as $ x \to - \infty$,
\be
\w(x) \sim Bi(\zeta )+ i Ai (\zeta ) \ \mbox { as } \ x \to + \infty \, ,
\ee
where $Ai$ and $Bi$ are the two Airy's functions and $
\zeta = -F^{1/3}x - F^{2/3} E $.

Let $\alpha >0$ be fixed then, assuming that $\psi_0$ is an entire vector for the translation group $\U^\lambda $, $\lambda \in \C$, we have that
\bee
\A (t) = \sum_{\mbox {\small resonances }E : -\alpha /2 \le \Im E <0 } C_E e^{-itE/\hbar } + r(t) \, ,  \label {F23}
\eee
where $
|r(t)| \le C e^{-t(\alpha + \epsilon )/2} $ for some $\epsilon >0$ and for some positive constant $C$ independent of $t >0$. \ In order to compute the numerical coefficients $C_E$ we have to consider the complex eigenvalues of the adjoint operator ${H^\lambda}^\star$. \ Since the restriction of $V(x)$ to the outer interval $\R \setminus I_2$ can be regarded as the restriction to this interval of the entire function $V(z)=-Fz$, $z \in \C$, such that $V(\bar z)=\overline {V(z)}$, then ${H^\lambda}^\star = {H^{\bar \lambda}}$. \ Hence $E$ is a discrete eigenvalue of $H^\lambda$ if, and only if, $\bar E$ is a discrete eigenvalue of the adjoint ${H^{\bar \lambda}}$ and the solution $v^\lambda \in L^2 (\R )$ to the equation $(H^{\bar \lambda} - \bar E )v^\lambda =0$ is given by $v^\lambda = \U^{\bar \lambda } v$ where $v\notin L^2 (\R )$ is the solution to the equation $(H-\bar E) v=0$ with the the outgoing Siegert's conditions $v(x) \sim 0$, as $ x \to - \infty$, and
\bee
 w(x) \sim Bi(\zeta )- i Ai (\zeta ) \ \mbox { as } \ x \to + \infty \, .
\eee
Eventually, let $\delta $ be such that $\delta >\alpha /2F$, then 
\bee
C_E 
= \frac {\langle \psi_{0}^{-i\delta },\w^{i \delta} \rangle \langle v^{i \delta} , \psi_{0}^{i\delta }  \rangle }{\langle v^{i \delta} , \w^{i \delta} \rangle}\, , \ \mbox { where } \ \psi_0^\lambda = \U^\lambda \psi_0 \, .\label {F26}
\eee
%
%\bibliography{bibliografia}
\printbibliography

%\printbibliography[heading=subbibliography, title={Methods References}]

%\end {refsection}

\bigskip

{\bf Data availability statement.} \ All data generated or analyzed during this study are included in this published article.

\bigskip

{\bf Competing interest.} \ The author declares no competing interests.

\end {document}